\documentclass[10pt]{article}
\usepackage{graphicx}
\usepackage{amsmath,times}

\title{Self-Compression and Controllable Guidance
of Multi-Millijoule Femtosecond Laser Pulses
}

\bigskip
\author{I. G. Koprinkov$^1$\footnote{corresponding author, email: igk@tu-sofia.bg}, M. D. Todorov$^2$, M. E. Todorova$^3$, and T. P. Todorov$^1$\\
\small\it $^1$Department of Applied Physics, Technical University of Sofia, 1000 Sofia, Bulgaria,\\[-1.mm]
\small\it $^2$Faculty of Applied Mathematics and Informatics, Technical University of Sofia, 1000 Sofia, Bulgaria\\[-1.mm]
\small\it $^3$Integrated Technical College, Technical University of Sofia, 1000 Sofia, Bulgaria}
\date{}

\begin{document}

\maketitle
\bigskip
\thispagestyle{empty}

\begin{abstract}
Self-compression of multi-millijoule femtosecond laser pulses and dramatic increase of the peak intensity are found in pressurized helium and neon within a range of intensity in which the ionization modification of the material parameters by the pulse is negligible. The pulse propagation is studied by the (3+1)-dimensional nonlinear Schrodinger equation including basic lowest order optical processes -- diffraction, group velocity dispersion of second order, and Kerr nonlinearity of third order. Smooth and well controllable pulse propagation dynamics is found. Constructing of compressed pulse of controllable parameters at given space target point can be achieved by a proper chose of the pulse energy and/or gas pressure.
\end{abstract}

\section{Introduction}
The propagation of high-intensity femtosecond laser pulses in a bulk medium is a complex phenomenon accompanied by a dramatic spatiotemporal and spectral rearrangement of the pulse. This originates from the highly intensive nonlinear and dispersive interaction of the optical pulse with the medium. The rigorous description of the pulse propagation is based on the (3+1)D nonlinear Schrodinger equation (NLSE), in which diffraction, group velocity dispersion (GVD) of second order, and Kerr nonlinearity of third order form a basic set of optical processes of lowest order [1-6], which is universal for all media and parameters of the input pulse and belongs to the strongest effects acting on the pulse in a centrosymmetric neutral medium. Additional terms in the NLSE are required to describe some particular features or more extreme conditions in the pulse propagation as, {\it e.g.}, self-steepening, space-time focusing, non-instantaneous nonlinearity, higher order dispersion, {\it etc.} [7-9]. Higher orders of nonlinearity and ionization must also be considered for pulses creating significant ionization [10-12]. While the femtosecond pulse behavior strongly depends on the material and pulse parameters, some general simple understandings were widely accepted. In this way, simultaneous collapse of the pulse in space and time and formation of a spatiotemporal soliton is predicted at negative GVD \cite{1}, while, in the more common case of positive GVD, the theoretical and experimental studies predict time broadening and splitting of the pulse [2-5,7,8,10]. In a single study, initial pulse shortening is found while not supported by experimental evidences, instead, splitting of the femtosecond pulse is shown \cite{6}.

Self-compression (SC), before splitting, of high-intensity femtosecond laser pulses in positive GVD has been discovered in various types of media -- pure atomic and molecular gases [13-15] and fused silica \cite{16}. This allows recognizing the SC as a new phenomenon -- a kind of a time focusing, or the temporal counterpart of the usual space domain self-focusing (SF) \cite{14}. The mechanism of both phenomena is different but, as will be shown below, the SC is closely related to the SF. The generation of compressed (almost five times in magnitude) pulses in the experiments in gaseous media is accompanied by a strong increase of the peak intensity, an improvement of the spatiotemporal pulse shape, and a stable propagation of the compressed pulse along a distance that is many times longer than the characteristic length of the strongest pulse rearrangement factor, in this case -- the nonlinearity \cite{14}. The time-spectral studies show that the compressed pulses are transform-limited \cite{17}. SC of high-intensity femtosecond laser pulses was also observed in air \cite{18}, and in ionized noble gasses [19-22]. Recently, SC was also reported in BK-7 glass using negatively chirped pulses \cite{23}. Generation of long light filament and twofold pulse compression has been observed experimentally in air \cite{18}. Substantially stronger pulse compression has been achieved in ionized noble gasses [19-22]. Generation of few-cycle pulses through filamentation is achieved experimentally in argon-filled gas cell \cite{20}. Generation of single-cycle pulse is predicted numerically in argon gas with pressure gradient \cite{21}. Grate attention has been paid in the study of the femtosecond pulse splitting. From the other side, while the SC is much more interesting for practical use, its exact ``physical mechanism ... is still not fully understood'' \cite{22}. Recently, we have shown for the first time that SC and intensity gain of high intensity femtosecond laser pulses, before the pulse splitting, can be achieved in a positive GVD medium based on basic set of the lowest order optical processes, diffraction, group velocity dispersion of second order and Kerr nonlinearity of third order, providing the GVD of the medium is low enough \cite{24}. The physical mechanism of SC in this case was found and referred to as SC in low dispersion regime. The numerical simulations were performed for pressurized argon assuming sub-millijoule femtosecond laser pulses so as to avoid the effect of the ionization on the pulse propagation. The lack of ionization substantially simplifies the pulse propagation dynamics. To scale up the above method to higher energy/intensity pulses, media of higher ionization potential have to be used. Here we demonstrate SC of multi-millijoule femtosecond laser pulses in pressurized helium and neon. The pulse compression is accompanied by a dramatic increase of the peak intensity at almost no loss of energy. The pressurized helium and neon possess the lowest specific GVD among the known media, which provides an increased degree of SC and peak intensity gain in comparison with argon. The dynamics of the optical pulse at lack of ionization is relatively smooth and it allows a reliable control on the pulse propagation and constructing a pulse of well predictable parameters in given space target point. The degree of SC and intensity gain, as well as the space positions of maximal SC and intensity gain of the pulse can be controlled by the pulse energy and/or gas number density/pressure.

\section{Physical Model}
We consider a physical model of the high-intensity femtosecond pulse propagation in a centrosymmetric nonlinear medium, which includes the basic set of lowest order optical processes, diffraction, GVD of second order and Kerr nonlinearity of third order. The NLSE (in moving frame)
\begin{equation}
\label{eq1}
\frac{\partial {\tilde E}}{\partial z} - \frac{\rm i}{2 k} \nabla^2_{\perp} {\tilde E} + \frac{{\rm i} \beta_2}{2} \frac{\partial^2 {\tilde E}}{\partial \tau^2} - \frac{{\rm i} k n_2}{n_0} |{\tilde E}|^2 {\tilde E}=0
\end{equation}
for the complex field amplitude ${\tilde E}(r,z,\tau)$  of a pulse propagating along the $z$-direction, including the above mentioned physical processes, is used as a propagation equation within that model. Here, $k=2\pi/\lambda$ is the magnitude of the wave-vector, $n$ is the linear refractive index of the medium at the central wavelength $\lambda$ of the pulse, $\nabla_\bot^2$ is the transversal Laplacian, $\beta_2$ is the GVD, and $n_2$ is the nonlinear refractive index. The field is normalized such that $I=|{\tilde E}|^2$ is the intensity.  Neon and helium are used as nonlinear media. The nonlinear refractive indexes of neon and helium, $n_2({\rm Ne})=7.8\times10^{-21}$ Pcm$^2$/W and $n_2({\rm He})=4.2\times10^{-21}$ Pcm$^2$/W, have been determined from a comparative study of the relative nonlinear refractive indexes of noble gas media \cite{25}, taking the nonlinear refractive index of Ar, $n_2({\rm Ar})=9.8\times10^{-20}$ Pcm$^2$/W, as a reference value. The GVD of neon and helium has been determined from the Sellmeier equation \cite{26} and found to be $\beta_2({\rm Ne})=0.02$ Pfs$^2$/cm  and $\beta_2({\rm He})=0.01$ Pfs$^2$/cm. In the above expressions $P$ is the pressure in atm. The initial pulse is a linearly polarized chirp-free Gaussian ${\tilde E}(r,z=0,\tau)=E_0 \exp(-r^2/2r_0^2-\tau^2/2\tau_0^2)$ in space and time having axial symmetry and 100 fs time duration (full width at half maximum (FWHM)) of the intensity profile. The central wavelength of the pulse is tuned at 800 nm.

The proper initial conditions are important in order to remain within the range of validity of the present model. For such a purpose, special attention is paid to the ionization. The ionization strongly and non-instantaneously modifies the material parameters and thus, the pulse propagation. Also, it is highly nonlinear process, which means that it is very sensitive to the changes of the intensity/field strength and its control is very difficult in practical means, more over -- the pulse intensity strongly changes during the propagation. The avoiding the ionization of medium can be achieved by a suitable choice of the input pulse energy and/or the gas number density/pressure so that the peak intensity, raised by the SF, does not surpass the level at which the ionization causes substantial modification of the material parameters. For instance, the plasma contribution to the refractive index should be kept negligible in comparison to the respective contribution of the cubic nonlinear Kerr effect within the whole range of intensity variation. Avoiding the ionization ensures a relatively simple, smooth and, thus, well controllable pulse propagation dynamics.

The determination of the absolute value of the ionization rate and, in this way, the plasma density created by the pulse itself is difficult task. The description of the ionization of atoms and simple molecules by the general ionization theories is not quite satisfactory \cite{27}. The ionization rate and the plasma number density predicted, {\it e.g.}, by Perelomov-Popov-Terent'ev (PPT) theory \cite{28}, which appears to be among the most successful ionization theories \cite{27}, closely matches the trend of the experimental data over a wide range of intensities -- from multi-photon to tunnel ionization regimes. However, the agreement between the experimental data and the theoretically predicted plasma number density, based on the PPT theory, is still within a factor of two. The discrepancy with the other ionization theories may reach orders of magnitude \cite{27}. From the other side, precision measurements of the ionization of helium \cite{29} as well as of the other noble gases \cite{27} are performed over a wide counting range. Unfortunately, only the relative value of the ion signal has been linked to the absolute value of the peak intensity of the pulses in these experiments [27,29], whereas, due to the ion detection technique, the absolute value of the plasma density remains undetermined. To account for the latter, we require at least one absolute reference point to calibrate the relative plasma number density-peak intensity relationships found from these experimental studies. Then, the precise relative dependence can be used to determine the respective absolute value dependence.  To find such a reference point, the following approach will be applied. The intensity of complete ionization of neon and helium atoms will be determined using the simple barrier suppression ionization (BSI) model \cite{30}. Within the BSI model, the threshold intensity $I_{\rm BSI}$ at which the optical field suppresses the Coulomb potential in an atom so as to allow a bound electron to escape freely above the potential barrier (above barrier ionization) without tunneling is (in atomic units)
\begin{equation}
\label{eq2}
I_{\rm BSI}=\frac{I_P^4}{16Z^4},
\end{equation}                                                                         		                                                 where $I_P$  is the ionization potential of the initial atom/ion, and $Z$ is the charge of the created ion. At the ionization potential of neon, $I_P=21.56$ eV, and helium, $I_P=24.58$ eV, the intensity threshold for complete ionization of the medium is found to be $I_{\rm BSI}({\rm Ne})=0.86\times10^{15}$ W/cm$^2$ and $I_{\rm BSI}({\rm He})=1.46\times10^{15}$ W/cm$^2$, respectively. The BSI intensity of helium atom so determined is relatively close to the experimentally determined value of $0.8\times10^{15}$ W/cm$^2$, beyond which indication of saturation of the ionization can be distinguished  (the slow $I^{3/2}$ increase of the experimentally measured ion yield beyond the sited intensity is attributed to the expansion of the Gaussian focal volume) \cite{29}. The intensity of the barrier suppression ionization, $I_{\rm BSI}$, corresponds to complete ionization of the medium, and, at $I=I_{\rm BSI}$, the plasma number density will be equal to the number density of the initial neutral gas (the plasma recombination will be neglected on the time scale comparable to the pulse duration of 100 fs). The latter is, in fact, the reference point that will be used here in order to find the relationship between the absolute values of the plasma number density and the peak intensity of the pulse. Using such a relationship allows to account for the intensity at which the plasma contribution to the refractive index is still negligible.

According to the above considerations, the electron/plasma number density $N_e$ at $I=I_{\rm BSI}$ will be equal to the initial number density of the neutral gas, $N_{no}$. We will take tentatively 6 orders of magnitude lower plasma number density than $N_{no}$, and the intensity at which such plasma number density is created will be determined approximately from the respective experimental curves, Figure 1 of Ref.\cite{29} and Figure 5 of Ref.\cite{27} for He, and Figure 4 of Ref.\cite{27} for Ne. From the sited figures, the intensity corresponding to such a plasma density was found to be about $I=2\times 10^{14}$ W/cm$^2$ and $I=1\times 10^{14}$ W/cm$^2$ for helium and neon, respectively. At the intensity so obtained and the respective plasma number density, the magnitude of, otherwise, negative plasma contribution to the refractive index, $\Delta n_{pl}\approx\omega_{pl}^2/2\omega^2=4\pi e^2 N_e/2m\omega^2$ \cite{31}, will be compared to the respective Kerr effect contribution to the refractive index, $n_{nl}=n_2|E|^2$. This is justified taking into account that the latter quantities determine the magnitude of the contributions of these effects to the respective terms, $k(\omega_{pl}^2/\omega^2)E$ and $kn_2|E|^2E$, in the NLSE \cite{12}. For 30 atm gas pressure of helium (the maximal gas pressure used here), the plasma contribution to the refractive index at $I=2\times 10^{14}$ W/cm$^2$ ({\it i.e.}, assuming, as sited above, 6 orders of magnitude lower plasma density than $N_{no}$) is $n_{pl}\approx \omega_{pl}^2/2\omega^2=2.15\times 10^{-7}$ while the Kerr effect contribution to the refractive index at the same intensity is $n_{nl}=n_2|E|^2=2.52\times 10^{-5}$. The same quantities for 17 atm pressure (around which most of the simulations are performed here) are $n_{pl}=1.22\times 10^{-7}$ and $n_{nl}=1.43\times 10^{-5}$ for the case of helium at $I=2\times 10^{14}$ W/cm$^2$ intensity, and $n_{pl}=1.22\times 10^{-7}$ and $n_{nl}=1.33\times 10^{-5}$ for the case of neon at $I=1\times 10^{14}$ W/cm$^2$ intensity. As can be seen, the plasma contribution to the modification of the refractive index is more than 2 orders of magnitude lower than the respective Kerr effect contribution and neglecting the plasma contribution seems justified for neon and helium at intensities not exceeding $I=1\times 10^{14}$ W/cm$^2$ and $I=2\times 10^{14}$ W/cm$^2$, respectively.

Within the above mentioned intensity range we are still in perturbative regime what concerns the neon and the helium atoms. The relation between two successive terms of the nonlinear polarization for bound-bound transitions is given by the following approximate relation \cite{32} $a_{bb}=\chi^{(n+2)} E^{n+2}/\chi^{(n)}E^n\approx(eE_aa_b/\hbar\Delta\omega)^2$ (for an isotropic medium the even order nonlinearities are zero), where $E_a$ is the time dependent field amplitude, $a_b$ is the Bohr radius, and $\hbar\Delta\omega=\hbar\omega_{eg}-\hbar\omega$ is the energy detuning of the laser photon energy $\hbar\omega$ (at the carrier frequency of the pulse) from the atomic resonance of energy $\hbar\omega_{eg}$ between the nearest exited atomic state (coupled to the ground state by a nonzero dipole moment) and the ground state. For the helium atom, the lowest excited level with nonzero dipole moment to the singlet ground state $1s$ is the singlet $2p$ state, lying about 21.2 eV above the ground state. At the field strength corresponding to the maximal intensity of $I=2\times10^{14}$ W/cm$^2$, the relation between the leading cubic term and the next quintic term is about $a_{bb}\approx10^{-2}$, or the truncation of the nonlinearity after the cubic term seems justified within the present considerations. One approach to control the peak intensity below the sited levels is to change the gas pressure while keeping the input energy fixed for given study. This is also a convenient approach for a real experiment.

In view of the above considerations, the propagation equation (\ref{eq1}) (which does not include irreversible losses and non-instantaneous processes) is appropriate to describe the propagation of femtosecond pulses within the conditions in which the present model is valid. The lack of substantial losses allows imposing energy conservation condition \cite{24}
\begin{equation}
\label{eq3}
W(z)=\int\limits_{S_{\infty}} \int\limits_{-\infty}^{+\infty} |{\tilde E}(r,z,\tau)|^2 {\rm d}s {\rm d}\tau = W_0 ={\rm const},
\end{equation}
where $W_0$ is the input pulse energy, $W(z)$ is the pulse energy along the propagation distance, and the integration is taken over the transversal cross-section $s$ and the local time $\tau$ of the pulse. The latter also represents a normalization condition for the complex field envelope $\tilde E$ along the propagation distance $z$. Applying the energy normalization condition for the field envelope allows to perform absolute value calculations.

\section{Results and Discussion}
The NLSE (\ref{eq1}) is solved numerically in a polar coordinate system (assuming axial symmetry) by using a split-step method similar to that applied in \cite{24}. Nonlinear media of highest ionization potentials~-- pressurized neon and helium, are considered in this study. For better comparison of the pulse compression capabilities of neon and helium, completely identical initial conditions for the optical pulse and the medium have been taken for the initial numerical simulations in both media. Our simulations show that at $W_0=2$ mJ input pulse energy and $P=17$ atm pressure of neon or helium, the maximal peak intensity does not surpass $I=1\times10^{14}$ W/cm$^2$ in neon and $I=2\times10^{14}$ W/cm$^2$ in helium along the whole propagation distance considered here. Consequently, at such initial conditions, the parameters of the pulse fall well within the range of validity of the present model.

The SC in low dispersion regime is inherently accompanied by a gain of the peak intensity of the pulse. That is why, two parameters have main importance here~ -- the magnitude of the peak intensity gain $k_{\rm IG}$ and the magnitude of the SC, $k_{SC}$. The peak intensity gain will be defined as the relation between the peak intensity of the pulse at given space point $z$, $I_0(z)$, and the peak intensity of the input pulse, $I_0(z=0)$, {\it i.e.}, $k_{\rm IG}(z)=I_0(z)/I_0(z=0)$. The magnitude of the SC will be defined as the relation between the time duration of the pulse (FWHM) at given space point $z$, $\tau_0(z)$, and the time duration of the input pulse $\tau_0(z=0)$, {\it i.e.}, $k_{\rm SC}(z)=\tau_0(z=0)/\tau_0(z)$.

The evolution of the transversal pulse width $2r_0$ (FWHM at $\tau=0$ plane), the normalized peak intensity $|E_0(r=0,z,\tau=0)/E_0(r=0,z=0,\tau=0)$ ({\it i.e.}, the peak intensity gain), and the pulse duration $\tau_0$ (FWHM) of the intensity profile along the propagation distance $z$ at 2 mJ input pulse energy and 17 atm pressure are shown in Figures 1(a),(b),(c) for neon and helium. The pulse shows similar behavior in both media, which is also close to the qualitative behavior of the pulse in the case of argon \cite{24}. The pulse rearrangement begins with a confinement of the pulse in the transversal direction due to the SF, Figure 1(a). The minimal width of the pulse in the transversal direction is reached at $z=z_{\rm SF}=0.653$ m in neon and $z=z_{\rm SF}=1.101$ m in helium. The SF confines the peripheral energy of the pulse toward the longitudinal axis, which leads to an increase of the peak intensity, Figure 1(b). The space positions at which the peak intensity reaches its maximal value is $z_{\rm IG}=0.642$ m in neon and $z_{\rm IG}=1.089$ m in helium. The corresponding maximal peak intensity gain at these positions in neon and helium is approximately $k_{\rm IG}(z=z_{\rm IG})=29$ and $k_{\rm IG}(z=z_{\rm IG})=48$ times, respectively. The most interesting and nonintuitive result is that the pulse undergoes SC in time (\emph{before splitting}) while propagating along a positive GVD medium, Figure 1(c). The position of the maximal SC is $z_{\rm SC}=0.620$ m in neon and $z_{\rm SC}=1.067$ m in helium. The magnitude of the SC in these positions is $k_{\rm SC}(z=z_{\rm SC})=3.4$ times in neon and $k_{\rm SC}(z=z_{\rm SC})=4.4$  times in helium. The intensity gain at the position of maximal SC is approximately $k_{\rm IG}(z=z_{\rm SC})=24$ times in neon and $k_{\rm IG}(z=z_{\rm SC})=37$ times in helium. As can be seen, the positions of the maximal SC and intensity gain do not coincide but are relatively close. This means that within given propagation distance simultaneous time compression and intensity gain of the pulse can be achieved. In a ``chronological'' order, first appears the maximal SC at $z_{\rm SC}$, next occurs the maximal intensity gain at $z_{\rm IG}$, and finally ~-- the maximal self-focusing at $z_{SF}$. Once the pulse reaches the maximal SC at $z=z_{\rm SC}$, it begins broadening in time from  $z_{\rm SC}$ to $z_{\rm IG}$ and the magnitude of the SC reduces to $k_{\rm SC}(z=z_{\rm IG})=2.8$ times in neon and $k_{\rm SC}(z=z_{\rm IG})=4$ times in helium. At the same time, the intensity continues to rise until $z=z_{\rm IG}$ because the pulse continue to narrow in the transversal direction due to the SF while it reaches the position of maximal SF at $z=z_{\rm SF}$. The propagation of the pulse beyond $z_{\rm IG}$ leads to the usual pulse splitting [2-8,10,24]. Indication of the pulse splitting in Figure 1(b) is the rapid fall of the peak intensity at the central peak point of the pulse at $\tau=0$ beyond $z=z_{\rm IG}$.
\begin{figure}[h]
\begin{center}
\includegraphics[width=2.4in]{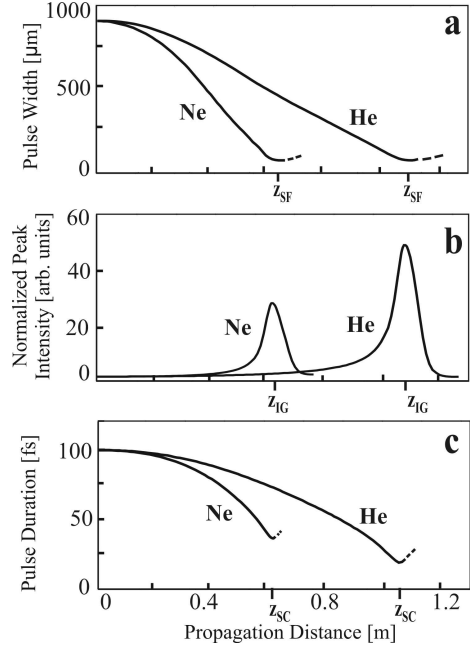}
\caption{Evolution of the transversal pulse width (FWHM) (a), normalized peak intensity (b), and pulse duration (FWHM) (c) versus the propagation distance. The positions where the respective quantities are not well defined because of the pulse splitting are indicated with dashed lines in the curves.}
\end{center}
\end{figure}

The comparative studies of the pulse compression in neon and helium clearly reveals a well expressed trend~ --- the pulse propagation dynamics develops more slowly in helium than in neon, while the magnitude of the pulse compression  and the intensity gain is substantially higher in helium than in neon at equal other conditions. This can be explained by the lower GVD of helium. The latter diminishes the effect of dispersion spreading of the pulse while it effectively compresses in time due to the SF. In fact, helium possesses the lowest specific GVD (at equal other conditions) and the highest ionization potential among the known media. That is why it appears to be the top ``speed'' medium within the present concept of propagation type compression of high intensity femtosecond laser pulses.

The optical pulse do not tend to form light filaments within the basic set of processes, but starts to diverge once it reaches the position of maximal SF, $z=z_{\rm SF}$, Figure 1(a). That is why, one may speculate that the generation of light filaments results from the defocusing effect of the above cubic nonlinearity, {\it e.g.}, $\chi^{(5)}$, and the ionization. These processes provide an instantaneous dynamical (due to the $\chi^{(5)}$-nonlinearity) and non-instantaneous (due to the ionization) balance of the $\chi^{(3)}$-SF and lead to formation of more or less stable light filaments [14,18,20]. For the experimental conditions considered here, the intensity raised by the $\chi^{(3)}$-SF does not reach high enough values so as to ``switch on'' the $\chi^{(5)}$  and the ionization defocusing, and the $\chi^{(3)}$-SF is restricted only by the diffraction and the pulse splitting due to the dispersion. The latter factors seem not to be sufficient to form relatively stable light filaments.

A general explanation of the SC in low dispersion regime has been found \cite{24} based on the energy conservation. For the case of localized electromagnetic fields (pulses), the energy conservation has a simple geometrical sense, expressed by Eq.(\ref{eq3}). According to it, the energy integral can be considered as a four-dimensional ``volume'' (two transversal directions over the cross-section of the pulse, a longitudinal direction -- the local time, and a ``vertical'' direction -- the intensity $|{\tilde E}|^2$), which must be conserved along the propagation coordinate $z$. An intensive rearrangement of the pulse is triggered by the SF coming from the nonlinear term. The rapid (relatively to the dispersive broadening) development of the SF is the main point in the pulse compression mechanism considered here. The SF confines the pulse radially from all transversal directions toward the longitudinal axis. This increases the peak intensity and further accelerates the pulse rearrangement. The increased peak intensity intensifies the self-phase modulation thus generating new spectral components -- the ``red''-frequency components are generated on the raising edge and the ``blue''-frequency components are generated on the falling edge of the pulse. In a positive GVD medium, this results in faster propagation of the fore edge and slower propagation of back edge of the pulse, which tends to cause pulse broadening. Due to the low dispersion of the rare gases neon and helium, the pulse would not substantially expand in time (within given propagation distance) if it is ruled solely on the dispersion. At the same time, the strong confinement of the pulse in the transversal direction due to the SF results in a strong increase of the pulse intensity, Figure 1(b). The latter leads to an effective, but \emph{real,} shortening of the pulse (based on the standard FWHM characterization of the pulse duration) before the pulse succeed to broaden substantially due to the dispersion. This is the mechanism of the pulse compression observed here. Such a concept of femtosecond pulse compression, formulated for the first time in \cite{24}, has been called \emph{self-compression in low dispersion regime.} The pulse compression in low dispersion regime is inherently accompanied by a strong increase of the peak intensity at almost no loss of energy, in accordance to the condition of conservation of the pulse energy. This is the main advantage of the present pulse compression method. In fact, the pulse behavior is subject of competition between two processes acting in opposite directions -- the pulse compression due to the intensity gain as a result of the SF and the pulse broadening due to the dispersion. The final outcome depends on the relative rates of both trends. Pulse compression along given propagation distance can be achieved if the rate of pulse broadening due to dispersion remains lower that the rate of pulse shortening due to intensity gain forced by the SF. Otherwise, pulse broadening and splitting will be observed. This has been confirmed numerically \cite{24}, showing that if the GVD exceeds given value (at equal other conditions), only pulse broadening and splitting, but not SC, is observed. This means that the SF does not necessarily lead to pulse compression [33,34]. The pulse compression in low dispersion regime is substantially related with the low value of GVD and the efficiency of this mechanism increases when the GVD diminishes. This explains why the degree of SC and the intensity gain increases from argon \cite{24} toward neon and helium, Figure 1. It is important to underline that the above SC phenomenon is substantially (3+1)D effect that is based on the SF and cannot be observed in the (1+1)D case at positive GVD ({\it e.g.}, optical fibers), where the SF is not considered. The latter also shows the mutual relation between the different dimensions in the high-intensity femtosecond pulse dynamics.

Although the pulse collapses in all dimensions in space and time, the confinement mechanism of the pulse within the basic set of optical processes is still incomplete. This is because confinement ``force'' in this case, {\it i.e.}, the SF, exists only in the transversal space dimensions but not in the time dimension at positive GVD. The SC in the time dimension results from the retarded dispersion broadening of the pulse due to the low GVD of the medium. Consequently, the stable pulse propagation in the positive GVD medium, as has been observed experimentally \cite{14}, requires additional process(es) to stabilize the optical pulse.

The present mechanism of pulse compression is based on the geometrical rearrangement of the pulse, in which the nonlinearity by means of the SF plays a leading active role, whereas the GVD, while it is also very important, has a secondary role, {\it i.e.}, to promote or to hamper the SC. That is why, such kind of SC is expected to work not only for initially chirp-free pulses, but for (regularly) chirped pulses and pulses having complicated, even irregular, frequency distortions. In the latter cases, the efficiency of such SC mechanism is subject of additional studies. The specified \emph{basic set} of optical processes is universal for all centrosymmetric media and experimental conditions. In particular, the basic lowest order processes are common for cases without or with ionization. That is why, the mechanism of SC described here should also play important role for the cases with substantial ionization of the medium. To what extend the proposed here mechanism of SC plays important role for the cases beyond the \emph{basic set,} besides on the particular experimental conditions, depends also on existing of other, if any, accompanying process(es) of comparatively strong action as those from the basic set. Among such additional processes, the most important are expected to be the ionization and above-cubic nonlinearities \cite{14}. The contribution of the ionization and the above-cubic nonlinearities (in the case of higher intensity pulses) to the present mechanism, as well as the efficiency of the latter, is subject of additional studies. No principle limitations seem to exist for the present mechanism of pulse compression to work for much shorter pulses then the ones used here but more advanced propagation equation \cite{32} have to be used in that case.

A kind of time shortening can also be found within the pulse splitting phenomenon, even if the initial pulse has newer reached shortening but broadens continuously and finally -- splits. At given stage of development of the split pulse, each of two sub-pulses, in which the initial pulse splits, may become shorter than the initial pulse. The split pulse, however, is usually considered as deteriorated pulse, having reduced energy and peak intensity in each subpulse, and such a kind of compression seems useless for the practical application. The SC before splitting, as it is observed here, must be clearly distinguished from the derivation of shorter subpulses within the pulse splitting phenomenon.
\begin{figure}[h]
\begin{center}
\includegraphics[width=3.8in]{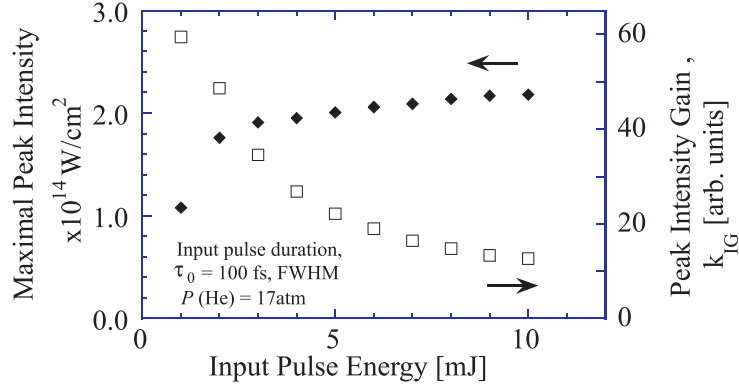}
\caption{Dependence of the maximal peak intensity and the respective peak intensity gain versus the input energy of the pulse at fixed   pressure of helium.}
\end{center}
\end{figure}
\begin{figure}[h]
\begin{center}
\includegraphics[width=3.3in]{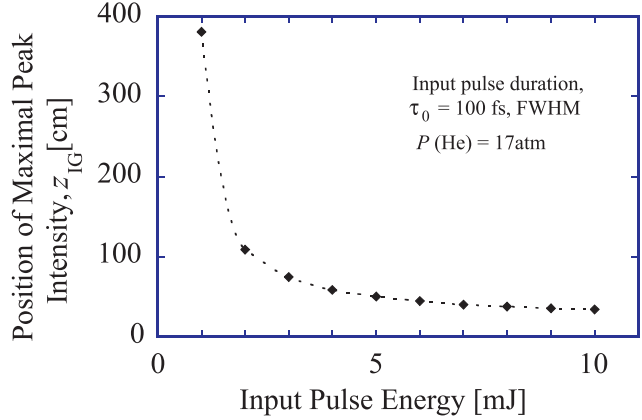}
\caption{Dependence of the space position $z_{\rm IG}$, where pulses of maximal intensity are generated, versus the input energy of the pulse at fixed 17 atm pressure of helium.}
\end{center}
\end{figure}

For comparison with the results for 2 mJ energy pulses, compression of higher energy 5 mJ-pulses were also studied in neon and helium, keeping the other conditions same as above.  In this case, we found lower degree of the peak intensity gain, $k_{\rm IG}(z=z_{\rm IG})=13$ times in neon and $k_{\rm IG}(z=z_{\rm IG})=22$ times in helium, which results in lower degree of pulse compression, $k_{\rm SC}(z=z_{\rm SC})=2.3$ times in neon and $k_{\rm SC}(z=z_{\rm SC})=3.0$ times in helium. Whereas the initial peak intensity is higher for 5 mJ input energy, the maximal peak intensity at $z_{\rm IG}$ does not substantially increases when the pulse energy increases from 2 mJ to 5 mJ. Thus, the maximal peak intensity at $z_{\rm IG}$ for the case of 2 mJ input pulse energy is $1\times10^{14}$ W/cm$^2$ in neon and $1.9\times10^{14}$ W/cm$^2$ in helium, while for the case of 5mJ these values are $1.2\times10^{14}$ W/cm$^2$ and $2\times10^{14}$ W/cm$^2$, respectively. The pulse compression is also lower at higher input energy. Consequently, the simple increase of the pulse energy does not necessarily lead to increase of the SC and the intensity gain and a proper optimization of the parameters of the initial pulse and the medium is required in order to achieve the highest performance from the pulse rearrangement process. In view of the non-intuitive SC behavior of the high-intensity femtosecond pulses versus the input pulse energy and medium density, a more detailed study on that subject is required and will be presented below.

Whereas the propagation of high intensity femtosecond laser pulses in the heavier rare gases (argon, krypton and xenon) has been widely investigated [10,11,13-15,19-22,24], the lighter rare gases (neon and, especially -- helium) remain insufficiently explored as propagation media. From the other side, the above results show the superior self-compression and intensity gain capabilities of helium in comparison with that of neon and argon \cite{24} within the present pulse compression concept. That is why, the further studies will be focused namely on helium. In the following, the dependence of the main parameters of the pulse on the input pulse energy and helium pressure will be shown. The development of the maximal peak intensity at $z=z_{\rm IG}$ and the respective magnitude of the peak intensity gain versus the input pulse energy at fixed helium pressure, $P=17$ atm, are shown in Figure 2. As one can see, the absolute value of the peak intensity increases with the input pulse energy, while, at the same time, the magnitude of the peak intensity gain, $k_{\rm IG}$, decreases. The increase of the peak intensity with the pulse energy is more rapid in the low energy scale and tends to saturate at high energies. The space position, $z_{\rm IG}$, where a pulse of the maximal peak intensity is created, as a function of the same values of the input pulse energies, as in Figure 2, is shown in Figure 3. The space position $z_{\rm IG}$, where the pulse has maximal intensity gain, changes in a kind of reciprocal way with the pulse energy. At relatively low energy, say $W_0=1$ mJ, the position of the maximal peak intensity is very far from the entrance of the medium, and it rapidly approaches the entrance when the pulse energy increases, Figure 3.
\begin{figure}[h]
\begin{center}
\includegraphics[width=3.2in]{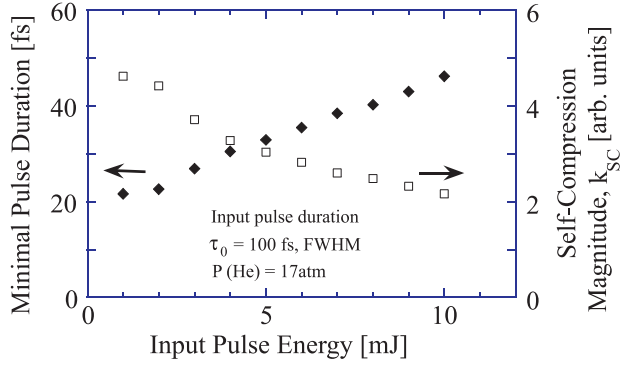}
\caption{Dependence of the minimal pulse duration and the respective coefficient of SC versus the input energy of the pulse at fixed 17 atm pressure of helium.}
\end{center}
\end{figure}
\begin{figure}[h]
\begin{center}
\includegraphics[width=3.2in]{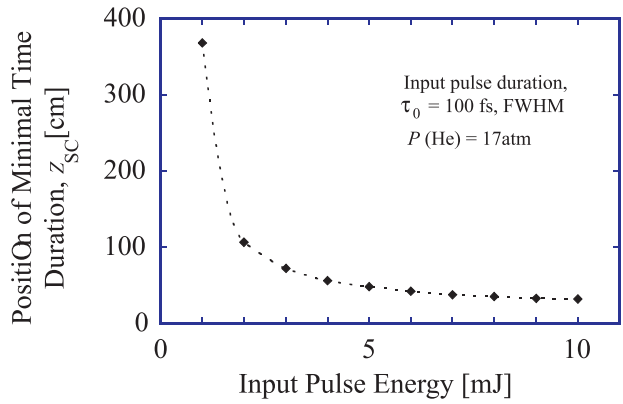}
\caption{Dependence of the space position $z_{\rm SC}$, where pulses of minimal pulse duration are generated, versus the input energy of the pulse at fixed 17 atm pressure of helium.}
\end{center}
\end{figure}

The dependence of the minimal pulse duration at $z=z_{\rm SC}$ and the respective magnitude of the pulse compression versus the input pulse energy at fixed helium pressure, $P=17$ atm, are shown in Figure 4. The space positions, $z_{\rm SC}$, where a pulse of minimal pulse duration is created, at the same input pulse energies, as in Figure 4,is shown in Figure 5. In contrast to the pulse intensity, Figure 2, the minimal pulse duration increases almost linearly with the input pulse energy. What is also interesting, the degree of the SC, $k_{\rm SC}$, decreases with the increase of the pulse energy. At the same time, the positions $z_{\rm SC}$ where maximally compressed pulses are created, change in a similar reciprocal way in Figure 5 as for the case of $z_{\rm IG}$ in Figure 3. At the lowest energy used, $W_0=1$ mJ, $z_{\rm SC}$ is very far from the entrance of the medium and it very rapidly moves toward the entrance when the energy increases, Figure 5.
\begin{figure}[h]
\begin{center}
\includegraphics[width=3.5in]{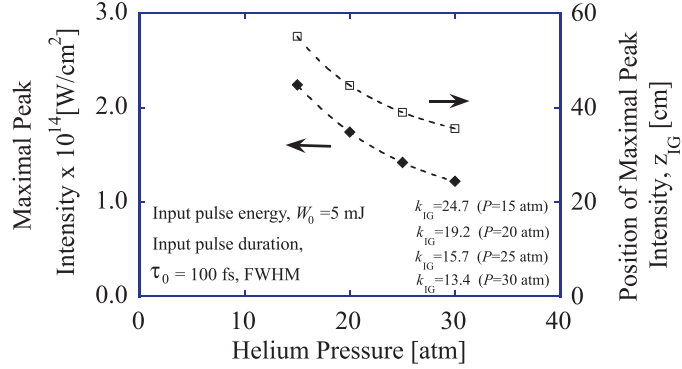}
\caption{Dependence of the maximal peak intensity and the position $z_{\rm IG}$, where such pulses are generated, versus the gas pressure of the medium at fixed $W_0=5$ mJ input energy of the pulse. The coefficients of the intensity gain $k_{\rm IG}$ at the respective values of gas pressure are also given in the figure.}
\end{center}
\end{figure}
\begin{figure}[h]
\begin{center}
\includegraphics[width=3.5in]{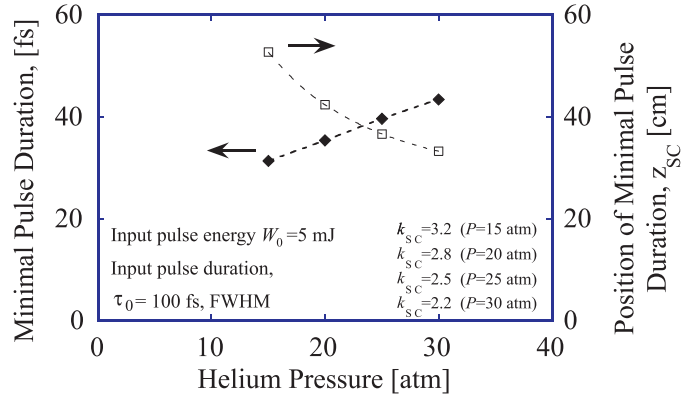}
\caption{Dependence of the minimal pulse duration and the position $z_{\rm IG}$, where such pulses are generated, versus the gas pressure of the medium at fixed $W_0=5$ mJ input energy of the pulse. The coefficients of the self-compression $k_{\rm SC}$ at respective values of gas pressure are also given in the figure.}
\end{center}
\end{figure}

The dependence of the maximal peak intensity at $z=z_{\rm IG}$ as well as the position $z_{\rm IG}$, where such pulses are generated, from the helium pressure at fixed input pulse energy of $W_0=5$ mJ is shown in Figure 6. The dependence of the minimal pulse duration at $z=z_{\rm SC}$  as well as the position $z_{\rm SC}$, where such pulses are generated, on the helium pressure at fixed input pulse energy of $W_0=5$ mJ is shown Figure 7. The coefficients of the intensity gain, $k_{\rm IG}$, and the SC, $k_{\rm SC}$, at the respective gas pressure, are also given in the figures. These studies show that the strongest degree of SC and intensity gain is achieved at lowest pressure, $P=15$ atm, which also means lowest value of the total GVD of the medium. The peak intensity at the lowest helium pressure in Figure 6, as well as at the highest energy pulses in Figure 3, surpasses the accepted here value of maximal intensity of $I=2\times 10^{14}$ W/cm$^2$ (for the case of helium) within about 10\%, and these results are shown for completeness of the study. Further reduction of the pressure results in peak intensity which is substantially beyond the specified value, and, for $W_0=5$ mJ pulses, the simulations were not performed below that pressure.

The evolution of the spatiotemporal pulse shape with the variations of gas pressure at fixed $W_0=5$ mJ input pulse energy has been also investigated. A ``time view'' of the spatiotemporal pulse shape taken at the position of maximal SC, $z=z_{\rm SC}$, is shown in Figure 8. The parameters of the predicted pulses at the specified space points are also given in the figure. These results demonstrate the rearrangement capabilities of the present pulse formation method at strong nonlinear interaction with the medium and low dispersion regime. If the medium ends around $z_{\rm SC}$  or within $z_{\rm SC}-z_{\rm IG}$ interval, a strongly compressed single femtosecond pulse of multi-millijoule energy and dramatically increased peak intensity will be generated. Input pulses of parameters used in these simulations can be generated by the present oscillator-chirped pulse amplifier Titan:Sapphire laser systems, which makes the results of this study suitable for practical applications.
\begin{figure}[h]
\begin{center}
\includegraphics[width=4.5in]{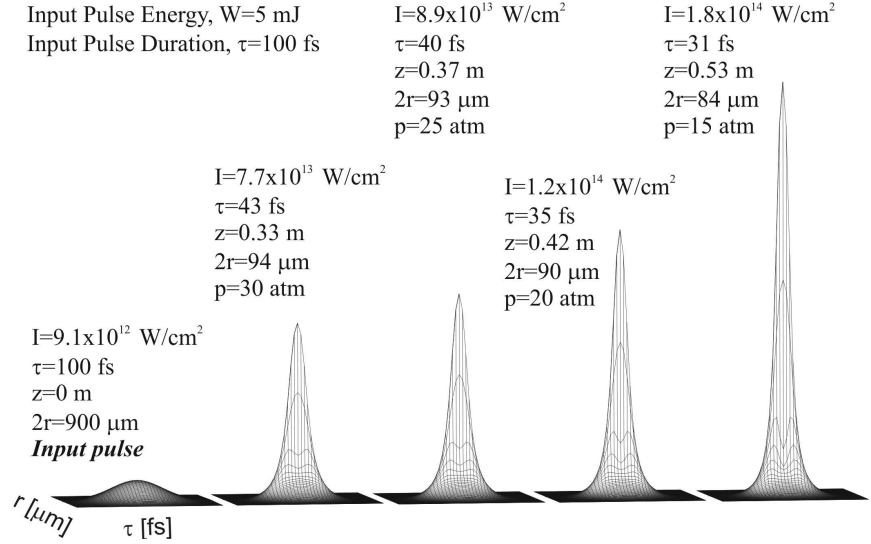}
\caption{Development of the spatiotemporal pulse shape (a ``time view'') and the respective parameters of the pulse taken at position of maximal time compression, $z=z_{\rm SC}$, versus the gas pressure of the medium.}
\end{center}
\end{figure}

All results found here directly follow from the solution of the NLSE. From the other side, the behavior of the pulse in regime of SC can be physically explained if we consider the interplay between the most active and important processes, in this case -- the SF and the dispersion. In the NLSE, they are described by the terms $ikn_2n_0^{-1}|{\tilde E}|^2 {\tilde E}$ (in the space domain) and $i(\beta_2/2)\partial^2{\tilde E}/\partial\tau^2$, respectively, Within such an approach, the slower development of the pulse compression in helium than in neon, Figure 1, can be explained with the lower nonlinear refractive index $n_2$ of helium and, at same peak intensity, smaller nonlinear term $ikn_2n_0^{-1}|{\tilde E}|^2 {\tilde E}$. This makes the process of SF in helium more gradual than in neon. From the other side, the higher value of intensity gain, and, as a result -- higher value of SC, in helium can be explained by the lower value of GVD of helium. While more slowly, the nonlinear term succeeds to SF the pulse in helium at almost same spot diameter, Figure 1(a). At the same time, the lower GVD in helium prevents the pulse from substantial time spreading. Thus, the common action of both terms ``squeeze'' stronger the pulse energy in the spatiotemporal domain, which results in a higher increase in the ``vertical'' direction, {\it i.e.}, intensity gain, and, thus -- stronger SC, Figure 1. The higher intensity gain and pulse compression at low input energy, Figure 2 and Figure 4, also looks naturally if it is considered within the interplay between above two terms. The low input energy for pulses of same initial pulse duration and same size in transversal direction means low input intensity and field amplitude. The latter also means low steepness of the pulse fronts and, thus, relatively low value of the time derivatives. As a final result, this diminishes the expanding rate due to the dispersion term $i(\beta_2/2)\partial^2{\tilde E}/\partial\tau^2$ in which, the second time derivative determines dynamically the magnitude of the expanding rate. This explains why at low input energy, the magnitude of the SC is higher. The above factors support the opposite trend with the increase of the input energy of the pulse. The higher input energy means higher input peak intensity and, consequently, higher value of the field time derivatives and higher magnitude of the dispersion term. The latter intensifies the process of the dispersion spreading and splitting, thus preventing the pulse to achieve higher levels of intensity gain and SC (in comparison with the lower energy case). The results in Figure 3 and Figure 5 also look transparent within the framework of such interplay. At low energy, but equal other conditions, the peak intensity and the field amplitude of the initial pulse is relatively low, which results in low rate of development of the self-focusing by the cubic term of the NLSE, $ikn_2n_0^{-1}|{\tilde E}|^2 {\tilde E}$. That is why, the position where the maximal SF, and, as a result, the maximal intensity gain and time compression occur is far from the entrance point of the medium, Figures 3,5. Increasing the input pulse energy, and thus, the input peak intensity, intensifies the SF and the pulse needs shorter distance, $z_{\rm IG}$ and $z_{\rm SC}$, to develop the maximal intensity gain and SC, respectively. Finally, the increased peak intensity gain, Figure 6, and, as a result, stronger SC (shorter pulse durations), Figure 7, with reduction of gas pressure can be explained by the reduced value of the GVD of the medium, and in this way -- reduced value of the whole dispersion term at lower pressure. This diminishes the dispersion spreading and promotes the intensity gain and the SC. Of course, the reduced pressure reduces also the (total) nonlinear refractive index of the medium, and, in this way, the whole nonlinear term. The reduction of the nonlinearity results in slower rate of development of the SF and extends distances $z_{\rm IG}$ and $z_{\rm SC}$ to generate pulses of maximal intensity and SC, see the data of $z_{\rm IG}$ and $z_{\rm SC}$ in Figures 6 and 7. The numerical simulations show that the net effect of the reduced dispersion dominates, and the final effect is increased pulse compression.

The existence of SC before the splitting has crucial importance for the application of this effect. For example, it allows, to certain degree, to overcome, or, at least, to relax the requirement for existing of stable soliton-like pulse propagation for some applications. The theoretical and experimental studies show that such pulses do not exist within the specified basic set of optical processes and experimental conditions, considered here. For many applications, however, it is not necessary to have a stable pulse along the whole propagation distance but only a pulse of specified parameters at given space point. Such point will be called space \emph{target point.} The generation of such pulses can be considered as a satisfactory alternative to the soliton generation at conditions not supporting solitons. In this way, the \emph{soliton concept} (where it is not applicable) can be replaced by a \emph{controllable guidance concept,} which we introduce here. The controllable guidance concept seems to be applicable in broad range of conditions, in which the pulse propagation dynamics can be unambiguously predicted. The absence of ionization in our case plays important role in the realization of such a concept because it ensures a smooth and irreversible (along the propagation distance of interest) pulse propagation dynamics. The latter consists in a single SC event, reaching maximal efficiency at $z_{\rm SC}$, followed by slight broadening of the pulse along a short distance (from $z_{\rm SC}$ to $z_{\rm IG}$), while the peak intensity continuously increases (reaching maximal value at $z=z_{\rm IG}$), and, finally, initiation/development of splitting of the pulse at/beyond $z=z_{\rm IG}$, Figure 1. Although the pulse evolves along the whole propagation distance, the SC mechanism together with the smooth pulse propagation dynamics facilitates the generation of a pulse of well predictable parameters at a desired space target point. It is based on the fact that the magnitude of the time compression and the intensity gain as well as the space position of the strongest time ``focusing'' $z_{\rm SC}$ and the strongest intensity gain $z_{\rm IG}$, can be controlled by the pulse ({\it e.g.}, energy), Figures  2-5, and the medium (density/pressure), Figures 6, 7, parameters. The pulse shape, Figure 8, also changes smoothly and could be a subject to control by the initial parameters. Such a control, however, has few shortcomings. At first, it is a passive kind of control and, once launched in the medium, the pulse is no more subject to external control. In the second place, such a control is not complete because the parameters of the compressed pulse cannot be controlled independently. In any case, however, the investigations toward the controllable guidance concept look perspective.
\begin{figure}[h]
\begin{center}
\includegraphics[width=1.5in]{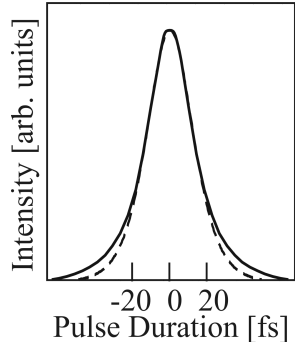}
\caption{Time $|E(r=0,z=z_{\rm SC},\tau)|^2$ intensity profiles of the tsunami pulse at the space position of maximal time compression,   at $P=17$ atm of helium. ${\rm Sech}^2$ time shape (dashed line) is also shown for comparison.}
\end{center}
\end{figure}

A very interesting feature of the pulse propagation can be distinguished from the results in Figures 1 and 8. As can be seen, the initial pulse has relatively long pulse duration and relatively low peak intensity. The parameters of the compressed pulse are most interesting for the practical applications around the points of maximal pulse compression, $z_{\rm SC}$, and/or the maximal intensity gain, $z_{\rm SC}$, which play role of target points. Far from given space target point, the pulse develops slowly and the peak intensity increases almost linearly with the propagation distance, Figure 1(b). Approaching the target point, the dynamical rearrangement of the pulse rapidly speeds up and its intensity increases in an exploding manner, Figure 1(b), whereas the pulse duration shortens gradually, Figure 1(c). This is also illustrated by Figure 8 if compare the initial pulse and some of the compressed pulse, say, at the lowest pressure -- $P=15$ atm. Such a behavior of an optical pulse closely resembles the behavior of the water tsunami wave, approaching the shore. That is why, such a compressed optical pulse whose intensity dramatically surges around the target points will be called \emph{optical tsunami pulse.} The time shape of the intensity profile of the compressed pulse at $z_{\rm SC}$ and $P=17$ atm of helium is shown in Figure 9. Although the input pulse has Gaussian temporal shape, the intensity time profile of the tsunami pulse is very close to ${\rm sech}^2$ pulse shape, also shown for comparison in Figure 9. The latter represents the temporal shape of the fundamental soliton of the (1+1)D NLSE in the case of medium of negative GVD \cite{35}. Insubstantial deviation of the shape of the tsunami pulse from that one of the fundamental soliton exists only around the wings of the pulse. Consequently, in accordance to the controllable guidance concept, a strongly compressed femtosecond optical tsunami pulse of huge intensity and soliton-like time intensity profile can be generated in given space point and the parameters of that pulse can be controlled within given limits by the input pulse and the medium parameters.

\section{Conclusions}
Self-compression of multi-millijoule femtosecond laser pulses has been found as a result of the interplay of the diffraction, group velocity dispersion of second order and Kerr nonlinearity of third order in lighter noble gases, neon and helium, at conditions of positive group velocity dispersion. These three processes are sufficient to trigger the self-compression at a low magnitude of positive group velocity dispersion of the medium. A comparative study of the self-compression in neon and helium has been performed and it confirms the importance of the low value of group velocity dispersion for the operation of the self-compression mechanism. The pressurized helium appears to be the best medium for compression the high energy femtosecond laser pulses within the pulse compression concept considered here. The parameters of the compressed pulse can be controlled within certain limits by the initial pulse and the medium parameters. A generation of compressed pulse of strongly increased intensity and soliton-like temporal profile in the vicinity of given space target point, called optical tsunami, is predicted.


\begin{thebibliography}{9}
\bibitem{17}
The initial pulses used in these experiments are not transform-limited while not, at the same time, regularly frequency modulated, or chirped. Instead, they possess complicated frequency distortions (sometimes, such pulses are also generally called -- ``chirped''), acquired within the process of amplification, that cannot be removed by purely dispersive pulse compressor (say, diffraction grating pulse compressor), as the regularly chirped pulses. That is why, SC of such pulses cannot be achieved by passing though a purely dispersive linear medium, which is a well known trivial effect, but results from a complex nonlinear pulse rearrangement process \cite{14}. This is especially clear from the pulse compression in solid bulk material, where the dispersion of the material remains constant and the SC results solely from an increase of the peak intensity, see Ref.\cite{16} of this work.
\end{thebibliography}
\end{document}